\documentclass[aps,prd,twocolumn,showpacs,showkeys,superscriptaddress]{revtex4}
\usepackage{amsmath}
\usepackage{graphicx}

\newcommand{\bastar}{\begin{eqnarray*}}
\newcommand{\eastar}{\end{eqnarray*}}
\newskip\humongous \humongous=0pt plus 1000pt minus 1000pt

\newif\ifdtup

\relax

\newcommand{\bea}{\begin{eqnarray}}
\newcommand{\eea}{\end{eqnarray}}
\newcommand{\pd}{\partial}
\newcommand{\A}{{\vec A}}

\newcommand{\nn}{\nonumber}

\newcommand{\mn}{{\mu\nu}}
\begin{document}
\title{Gravitationally Coupled Electroweak Monopole}
\bigskip
\author{Y. M. Cho}
\email{ymcho7@konkuk.ac.kr}
\affiliation{Administration Building 310-4, 
Konkuk University, Seoul 143-701, Korea}
\affiliation{School of Physics and Astronomy, 
Seoul National University,
Seoul 151-742, Korea}
\author{Kyoungtae Kimm}
\affiliation{Faculty of Liberal Education,
Seoul National University, Seoul 151-747, Korea}
\author{J. H. Yoon}
\affiliation{Department of Physics, College of Natural 
Sciences, Konkuk University, Seoul 143-701, Korea}

\begin{abstract}
We present a family of gravitationally coupled electroweak 
monopole solutions in Einstein-Weinberg-Salam theory. 
Our result confirms the existence of globally regular 
gravitating electroweak monopole which changes to 
the magnetically charged black hole as the Higgs vacuum 
value approaches to the Planck scale. Moreover, our 
solutions could provide a more accurate description of 
the monopole stars and magnetically charged black holes. 
\end{abstract}


\maketitle

Ever since Dirac has proposed the Dirac monopole generalizing 
the Maxwell's theory, the monopole has become an obsession 
theoretically and experimentally \cite{dirac}. After Dirac we have 
had the Wu-Yang monopole \cite{wu}, the 't Hooft-Polyakov 
monopole \cite{thooft}, the grand unification (Dokos-Tomaras) 
monopole \cite{dokos}, and the electroweak (Cho-Maison) 
monopole \cite{plb97,yang1,yang2}. But none of them except 
the electroweak monopole might become realistic enough 
to be discovered. 

Indeed the Dirac monopole in electrodynamics should 
transform to the electroweak monopole after the unification 
of the electromagnetic and weak interactions, and Wu-Yang 
monopole in QCD is supposed to make the monopole 
condensation to confine the color. Moreover, the 't Hooft-Polyakov 
monopole exists in an unphysical theory, and the grand unification 
monopole which could have existed at the grand unification 
scale probably has become completely irrelevant at present 
universe after the inflation. 

This makes the electroweak monopole the only realistic monopole
we could ever hope to detect, which has made the experimental 
confirmation of the electroweak monopole one of the most urgent 
issues in the standard model after the discovery of the Higgs particle 
at LHC. In fact the newest MoEDAL (``the magnificient seventh") 
detector at LHC is actively searching for the monopole \cite{moedal}.
But to detect the electroweak monopole at LHC, we have to ask 
the following questions.

First, does the electroweak monopole really exist? This, of course, 
is the fundamental question. As we know, the Dirac monopole in 
electrodynamics does not have to exist, because there is no reason 
why the electromagnetic U(1) gauge group has to be non-trivial. 
So we must know if the standard model predicts the monopole
or not. 

Fortunately, unlike the Dirac monopole, the electroweak 
monopole must exist. This is because the electromagnetic 
U(1) in the standard model is obtained by the linear 
combination of the U(1) subgroup of SU(2) and the hypercharge 
U(1), but it is well known that the U(1) subgroup of SU(2) 
is non-trivial. In this case the mathematical consistency 
requires the electromagnetic U(1) non-trivial, so that 
the electroweak monopole must exist if the standard model 
is correct \cite{epjb15,mpla16}. But this has to be confirmed 
by experiment. This makes the discovery of the monopole, 
not the Higgs particle, the final (and topological) test of 
the standard model. 

Second, what (if any) is the charactristic feature 
of the electroweak monopole which is different from 
the Dirac monopole? This is an important question 
for us to tell the  monopole (when discovered) is 
the Dirac monopole or the electroweak monopole. 
The charactristic difference is the magnetic charge.
The electroweak monopole has the magnetic charge 
twice bigger than that of the Dirac monopole. 
The magnetic charge of the Dirac monopole becomes 
a multiple of $2\pi/e$, because the period of 
the electromagnetic U(1) is $2\pi$. 

On the other hand, the magnetic charge of 
the electroweak monopole becomes a multiple 
of $4\pi/e$, because the period of electromagnetic 
U(1) in the standard model becomes 
$4\pi$ \cite{plb97,epjb15,mpla16}. The reason is
that this U(1) is (again) given by the linear 
combination of the U(1) subgroup of SU(2) and 
the hypercharge U(1), but the period of the U(1) 
subgroup of SU(2) is well known to be $4\pi$.  

Third, can we estimate the mass of the electroweak 
monopole? This is the most important question from 
the experimental point of view. There was no way to 
predict the mass of the Dirac monopole theoretically, 
which has made the search for the monopole a blind 
search in the dark room without any theoretical lead. 

Remarkably the mass of the electroweak monopole can 
be predicted. Of course, the Cho-Maison monopole has 
a singularity at the origin which makes the energy 
divergent \cite{plb97}. But there are ways to regularize 
the energy and estimate the mass, and all point consistently 
to 4 to 10 TeV \cite{epjb15,mpla16,ellis}. This, however, 
is tantalizing because the upgraded 14 TeV LHC can produce 
the electroweak monopole pairs only when the monopole 
mass becomes below 7 TeV. So we need a more accurate 
estimate of the monopole to see if LHC can actually 
produce the monopole. 

The purpose of this Letter is discuss how the gravitational 
interaction affects the electroweak monopole. We show that, 
when the gravity is turned on, the monopole becomes a globally
regular gravitating electroweak monopole which looks very much 
like the non-gravitating monopole, but turns to the magnetic 
black holes as the Higgs vacuum value approaches to the Planck 
scale. This confirms that the change of the monopole mass due 
to the gravitational interaction is negligible, which assures 
that the present LHC could produce the electroweak monopole. 

Before we discuss the modification of the monopole induced 
by the gravitation, we briefly review the non-gravitating 
electroweak monopole and explain how we can estimate 
the monopole mass. Consider the following effective Lagrangian 
of the standard model,  
\begin{gather}
{\cal L}_{eff} = -|{\cal D} _\mu \phi|^2
-\frac{\lambda}{2} \Big(\phi^2 -\frac{\mu^2}{\lambda}\Big)^2 
-\frac{1}{4} \vec F_\mn^2 -\frac{\epsilon(\phi)}{4} G_\mn^2,  \nn\\
{\cal D}_{\mu} \phi=\big(\partial_{\mu} 
-i\frac{g}{2} \vec \tau \cdot \vec A_{\mu} 
- i\frac{g'}{2}B_{\mu}\big) \phi,
\label{effl}
\end{gather}
where $\epsilon(\phi)$ is a positive dimensionless function 
of the Higgs doublet which approaches to unit asymptotically. 
Obviously when $\epsilon =1$, the Lagrangian reproduces 
the standard model. In general $\epsilon$ modifies 
the permeability of the hypercharge U(1) gauge field, 
but the effective Lagrangian still retains the SU(2) 
$\times$ U(1) gauge symmetry. 

When $\epsilon=1$, we can obtain the Cho-Maison monopole 
with the ansatz \cite{plb97}
\begin{gather}
\phi=\frac{1}{\sqrt 2}~\rho \xi,~~\rho = \rho(r),
~~\xi = i \left(\begin{array}{c}
\sin \theta/2 ~e^{-i\varphi} \\ -\cos \theta /2
\end{array}  \right), \nn\\
\A_\mu =\frac{1}{g} (f(r) - 1) \hat{r} \times \pd_\mu\hat{r}, \nn\\
B_\mu = -\frac{1}{g'} (1 - \cos\theta) \pd_\mu \varphi.
\label{cmans}
\end{gather}
Notice that $\A_\mu$ has the structure of the 't Hooft-Polyakov 
monopole, but $B_\mu$ has the structure of the Dirac monopole.
This tells that the Cho-Maison monopole is a hybrid between 
't Hooft-Polyakov and Dirac. 

The ansatz clearly shows that the U(1) point singularity 
in $B_\mu$ makes the energy of the Cho-Maison infinite, 
so that classically the monopole mass is undetermined. 
But, unlike the Dirac monopole, here we can estimate 
the mass of the electroweak monopole. A simplest way to do 
so is to realize that basically the monopole mass comes from 
the same mechanism which generates the weak boson mass, 
i.e., the Higgs mechanism, except that here the coupling 
becomes magnetic (i.e., $4\pi/e$) \cite{epjb15,mpla16}. 
This must be clear from (\ref{cmans}). So, from the dimensional 
reasoning the monopole mass $M$ should be of the order of 
$M\simeq (4\pi/e^2)\times M_W$, or roughly about 10 TeV.    

\begin{figure}
\includegraphics[width=8cm, height=4cm]{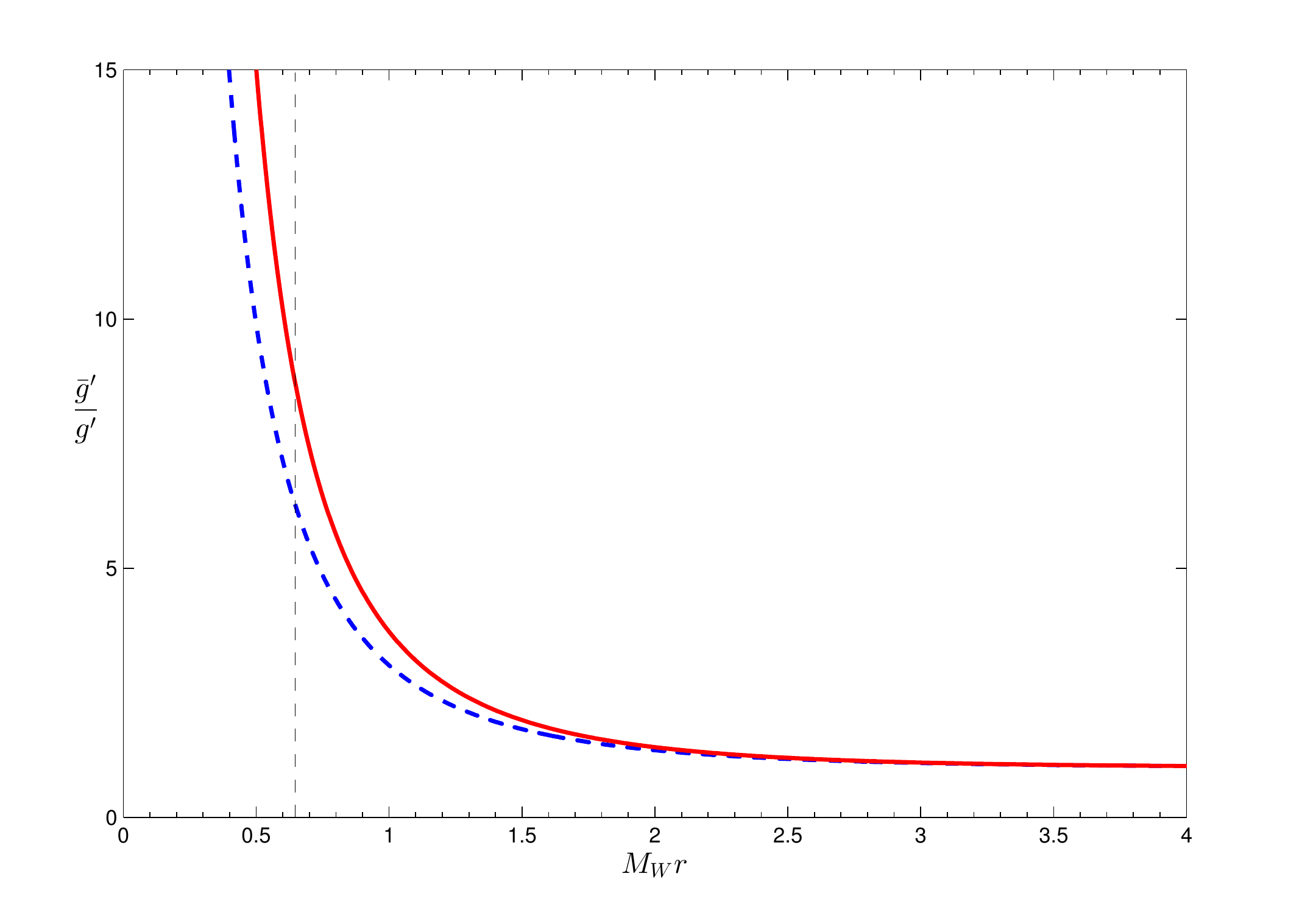}
\caption{\label{effg} The running coupling $\bar g'$ of 
the hypercharge U(1) induced by $\epsilon$. The dotted 
(blue) curve is obtained with $\epsilon=(\rho/\rho_0)^8$, 
and the solid (red) curve is obtained with $\epsilon$ 
proposed by Ellis et al. The vertical line indicates 
the Higgs mass scale.} 
\end{figure}   

A better way to estimate the mass is to notice that 
the Cho-Maison monopole energy consists of four parts, 
the SU(2) part, the hypercharge U(1) part, the Higgs 
kinetic part, and the Higgs potential part, but only 
the U(1) part is divergent \cite{plb97,epjb15}. Now, 
assuming that this divergent part can be regularized 
by the ultra-violet quantum correction, we can derive 
a constraint among the four parts which minimizes 
the monopole energy using the Derrick's theorem. From 
this we can deduce the monopole energy to be around 
3.96 TeV \cite{epjb15,mpla16}. 

Moreover, we can regularize the Cho-Maison monopole
introducing non-vacuum permeability $\epsilon$ which
can mimic the quantum correction. This is because, 
with the rescaling of $B_\mu$ to $B_\mu/g'$, $g'$ changes 
to $g' /\sqrt{\epsilon}$, so that $\epsilon$ changes 
the U(1) gauge coupling $g'$ to the ``running" coupling 
$\bar g'=g' /\sqrt{\epsilon}$. So, by making $\bar g'$ 
infinite (requiring $\epsilon$ vanishing) at the origin, 
we can regularize the monopole. For example, with 
$\epsilon=(\rho/\rho_0)^8$, the regularized monopole 
energy becomes 7.19 TeV \cite{epjb15,mpla16}.

The monopole energy, of course, depends on the functional 
form of $\epsilon$, so that we could change the monopole 
energy changing $\epsilon$. Recently Ellis et al. pointed 
out that $\epsilon=(\rho/\rho_0)^8$ is unrealistic because 
it makes the Higgs to two photon decay rate larger than 
the experimental value measured by LHC. Moreover, they 
have argued that the monopole mass can not be larger 
than 5.5 TeV if we choose a more realistic $\epsilon$ 
which reproduces the experimental value of the Higgs to 
two photon decay rate \cite{ellis}. The effective couplings 
induced by two different $\epsilon$ are shown in 
Fig. \ref{effg}.

Now we discuss the gravitational modification of the monopole. 
Intuitively, the gravitational modification is expected 
to be negligible. But there is the possibility that the gravitational 
attraction could change the monopole to a black hole and make 
the monopole mass arbitrary \cite{prd75}. We show that this 
happens only when the Higgs vacuum value approaches to 
the Planck scale. 
  
Consider the effective Lagrangian (\ref{effl}) minimally 
coupled to Einstein's theory. For the sake of simplicity, 
we focus on the static spherically symmetric solutions and 
assume the space-time metric to be
\begin{gather}
ds^2 = -N^2 (r) A(r) dt^2 + \frac{dr^2}{A(r)} 
+ r^2 (d^2\theta + \sin ^2 \theta d\varphi^2),  \nn\\
A(r)= 1 - \frac{2Gm (r)}{r}. 
\end{gather}
Now, with the ansatz (\ref{cmans}) the reduced 
Einstein-Weinberg-Salam action becomes  
\begin{gather}
S= \int \Big[\frac{1}{4\pi } \dot{m}  - A K - U \Big] N dr,  \nn\\
K = \frac{\dot{f}^2}{g^2 } +  \frac{r^2}{2} \dot{\rho}^2,  \nn\\
U = \frac{(1-f^2)^2}{2g^2 r^2}+ \frac{\lambda}{8}r^2 (\rho^2 
- \rho_0^2)^2 + \frac{ \epsilon(\rho)  }{2g'^2 r^2 }
+ \frac{1}{4} f^2 \rho^2,
\label{react}
\end{gather}
where the dot represents $d/dr$.

From this we have the following equations of motion 
\begin{gather}
\frac{\dot{N}}{N} = 8\pi G \frac{K}{r}, 
~~~\dot{m} = 4\pi   (A K + U), \nn\\
A \ddot{f} + \Big(\dot{A} + A \frac{\dot{N}}{N} \Big) \dot{f}
+ \frac{1-f^2}{r^2 }f - \frac{1}{4}g^2 \rho^2 f=0,\nn\\
A \ddot{\rho} + \Big( \frac{2A}{r} + \dot{A} 
+ A \frac{\dot{N}}{N}\Big) \dot{\rho} -\frac{f}{2r^2}\rho  \nn\\
-\frac{\lambda}{2} (\rho^2- \rho^2 _0)\rho 
-\frac{1}{2g'^2 r^4} \frac{d\epsilon(\rho)}{d \rho}= 0. 
\label{gmeq}
\end{gather} 
This has two limiting solutions. First, when gravitational 
field decouples (i.e., when $G\rightarrow 0$) we have 
the non-gravitating electroweak monopole solution \cite{epjb15}. 
Second, when $f=0$, $\rho=\rho_0$, and $\epsilon=1$, 
we have the magnetically charged Reissner-Nordstrom 
black hole solution with the magnetic charge 
$4\pi/e~(e=gg'/\sqrt{g^2+g'^2})$ \cite{prd75}. But in 
general the solution depends on three parameters
\begin{gather}
\alpha = \sqrt G \rho_0 = \rho_0/M_p,
~~~\beta = M_H/M_W,
\end{gather}
and the Weinberg angle $\theta_W$. 

\begin{figure}
\includegraphics[width=8cm, height=4cm]{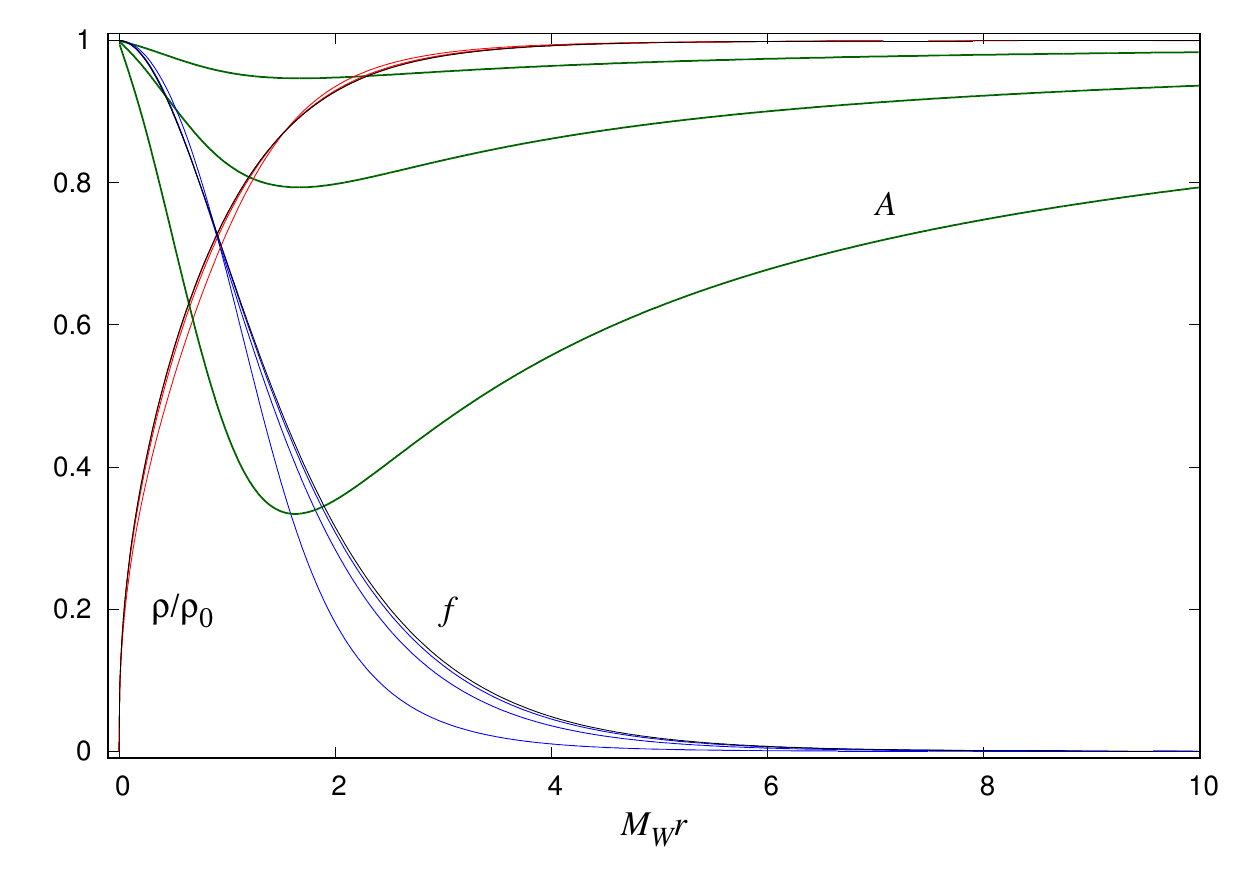}
\caption{\label{gmono} The W-boson $f$ (blue), Higgs field 
$\rho$ (red), and the metric function $A(r)=1-2Gm/r$ 
(green) profiles of the gravitating monopole solutions 
in the standard model. The black curves represent the finite 
energy solution in flat space-time, and the other three 
are obtained with $\alpha = \rho_0/M_P=$ 0.1, 0.2, and 
0.38.}
\end{figure}

Depending on the boundary conditions the entire solutions 
of (\ref{gmeq}) can be classified into two categories: 
the globally regular solutions and the black holes. Asymptotic 
flatness of the space-time requires that both $N(r)$ and 
$m(r)$ become constants at spatial infinity, which requires 
the following boundary conditions on the W-boson and Higgs 
field,
\begin{gather}
f(\infty) = 0,~~~ \rho(\infty) = \rho_0.
\label{bc1}
\end{gather}
We fix the scale of the time coordinate adopting 
\begin{gather}
N(\infty) =1.
\label{nbc}
\end{gather}
Notice that $m(\infty)$ which determines the total mass 
of the monopole is not constrained.

For the regular monopole solutions we require the regularity 
at the origin, 
\begin{gather}
f(0) =1,~~~\rho(0) =0,~~~m(0)=0.
\end{gather}
Now, choosing $\epsilon=(\rho/\rho_0)^8$ for simplicity,
we find that the solutions have the following expansions 
near $r=0$,
\begin{gather}
f(r) = 1- f_1 x^2+...,   \nn \\
\rho(r) = h_1 \rho_0 x^{\delta_1}+...,
~~~(\delta_1 =\frac{\sqrt{3}-1}{2})  \nn \\
m(r) =\frac{2\pi \alpha^2 h_1^2\delta_1^2}{GM_W \delta_2} 
x^{\delta_2}+...,~~~(\delta_2=\sqrt{3}),
\end{gather}
where $x=M_W r$. From this we can obtain the solutions 
by the standard shooting method with $f_1$ and $h_1$ as 
the shooting parameters. The result is shown in Fig. \ref{gmono}. 
Notice that, except for the metric, the gravitating monopole 
looks very much like the non-gravitating monopole. In particular, 
the size of the monopole remains roughly about $1/ M_W$. 

\begin{table}[htbp*]
\label{t}
\caption{\label{t} The numerical estimate of ADM mass 
of gravitating monopole with $\epsilon = (\rho/\rho_0)^8$ 
and physical value of $\beta$.}
\begin{center}
\begin{tabular}{|c|c|} \hline \hline
$ \alpha=\rho_0/M_P $ & ${\cal M}$ \\ \hline 
~0 (non-gravitating)~ & ~~~7.19~TeV~~~  \\  \hline
0.10  & ~~~7.15~TeV~~~  \\  \hline 
0.20  & ~~~6.97~TeV~~~  \\  \hline 
0.38  & ~~~6.34~TeV~~~  \\  \hline
$\alpha_\text{max}\simeq $ 0.39 & ~~~black hole~~~ \\
\hline\hline
\end{tabular}
\end{center}
\end{table}

To find the mass of the monopole, first note that with 
(\ref{gmeq}) we have 
\begin{gather}
m(r) = 4\pi   e^{P(r)} \int_0^r (K + U) e^{-P(r')}dr',  \\
P(r) = 8\pi G \int_r^\infty \frac{K}{r'} dr'.
\end{gather}
So, with (\ref{bc1}) we have the total mass given by
\begin{gather}
{\cal M} = m(\infty)  
=4\pi   \int_0^\infty  (K + U) e^{-P(r)}dr,
\end{gather}
which assures the positivity of the total mass. Moreover, 
this confirms that the gravitation reduces the monopole 
mass, which is expected from the attractive nature of 
the gravitational interaction \cite{vannieu}. In reality, 
of course, the gravitational modification of the monopole 
mass becomes negligible because $\alpha$ is very small 
(of $10^{-16}$). The $\alpha$-dependence of the mass is 
calculated numerically in Table \ref{t}. 

Obviously the mass depends on the permeability function
$\epsilon$. Here we have used $\epsilon=(\rho/\rho_0)^8$ 
for simplicity, but if we use the more realistic $\epsilon$ 
adopted by Ellis et al., the mass should become smaller 
than 5.5 TeV. From this we may conclude that the electroweak 
monopole mass can be 4 to 5.5 TeV. 

When the Higgs vacuum value approaches to the Planck mass 
and the monopole size becomes comparable to its Schwarzschild 
radius (roughly $GM_W/e^2$), however, the gravitational 
instability takes place and the solution turns to a black 
hole. So the globally regular monopole solutions can exist 
up to some maximal value $\alpha_\text{max}$. This implies 
that the grand unification monopole (with the huge Higgs
vacuum value) could be much more sensitive to the gravitational 
instability than the electroweak monopole, and more easily 
change to a black hole when coupled to the gravity.

Moreover, with $\epsilon(\rho(\infty))=1$, the long-range 
tail of the monopole is given by
\begin{gather}
m(r) = {\cal M} - \frac{2\pi}{e^2}\frac{1}{r} + ....
\end{gather}
This shows that (even for the globally regular solutions) 
the metric for the monopole becomes asymptotically 
Reissner-Nordstrom type, with ADM mass equal to ${\cal M}$ 
and the magnetic charge $Q_m =4\pi/e$. 

Our result shows that the generic feature of the gravitating 
electroweak monopole solution is quite similar to 
the gravitating 't Hooft-polyakov monopole solution obtained 
by Breitenlohner, Forgacs, and Maison \cite{bfm}. In fact 
mathematically our solution could be viewed as the electroweak 
generalization of their solution. From the physical point 
of view, however, they are totally different. Theirs is hypothetical, 
but ours describes the real monopole dressed by the physical 
W-boson and Higgs field.  

Clearly our solutions have the dyonic generalization, and 
have mathematically very interesting properties to study. 
Moreover, they must have important applications on the monopole 
stars and the magnetically charged black holes. The details 
of our solutions and their physical applications, in particular 
the dyonic generaliation and the comparison between our 
solutions and the Breitenlohner-Forgacs-Maison solutions, 
will be discussed in a separate paper \cite{cho}.

{\bf ACKNOWLEDGEMENT}

~~~The work is supported in part by the National Research 
Foundation of Korea funded by the Ministry of Education 
(Grants 2015-R1D1A1A0-1057578 and 2015-R1D1A1A0-1059407), 
and by Konkuk University.


\begin{thebibliography}{99}
\bibitem{dirac} P.A.M. Dirac, Proc. Roy. Soc. London, 
{\bf A133}, 60 (1931); Phys. Rev. {\bf 74}, 817 (1948).
\bibitem{wu} T.T. Wu and C.N. Yang, in {\it Properties of Matter 
under Unusual Conditions}, edited by H. Mark and S. Fernbach 
(Interscience, New York) 1969; Y.M. Cho, Phys. Rev. Lett. 
{\bf 44}, 1115 (1980); Phys. Lett. {\bf B115}, 125 (1982).
\bibitem{thooft} G. 't Hooft, Nucl. Phys. {\bf B79}, 276 (1974);
A.M. Polyakov, JETP Lett. {\bf 20}, 194 (1974); 
M. Prasad and C. Sommerfield, Phys. Rev. Lett. {\bf 35}, 
760 (1975).
\bibitem{dokos} C. Dokos and T. Tomaras, Phys. Rev. {\bf D21}, 2940 (1980).
\bibitem{plb97} Y.M. Cho and D. Maison, Phys. Lett. {\bf B391}, 
360 (1997); W.S. Bae and Y.M. Cho, JKPS {\bf 46}, 791 (2005).
\bibitem{yang1} Yisong Yang, Proc. Roy. Soc. London, {\bf A454}, 155 (1998). 
\bibitem{yang2} Yisong Yang, {\it Solitons in Field Theory 
and Nonlinear Analysis} (Springer Monographs in Mathematics), 
p. 322 (Springer-Verlag) 2001.
\bibitem{moedal} B. Acharya et al. (MoEDAL Collaboration), 
Int. J. Mod. Phys. {\bf A29}, 1430050 (2014); 
Y.M. Cho and J. Pinfold, Snowmass White paper, arXiv: [hep-ph] 1307.8390.
\bibitem{epjb15} Kyoungtae Kimm, J.H. Yoon, and Y.M. Cho, 
Eur. Phys. J. {\bf C75}, 67 (2015).
\bibitem{mpla16} Kyoungtae Kimm, J.H. Yoon, S.H. Oh, and Y.M. Cho, 
Mod. Phys. Lett. {\bf A31}, 1650053 (2016).
\bibitem{ellis} J. Ellis, N.E. Mavromatos, and T. You, 
Phys. Lett {\bf B756}, 29, (2016).
\bibitem{prd75} F.A. Bais and R.J. Russell, Phys. Rev. {\bf D11}, 2692 (1975);
Y.M. Cho and P.G.O. Freund, Phys. Rev. {\bf D12}, 1588 (1975).
\bibitem{vannieu} P. van Nieuwenhuizen, D. Wilkinson, and M. J. Perry, 
Phys. Rev. {\bf D13}, 778 (1976).
\bibitem{bfm} P. Breitenlohner, P. Forgacs, and  D. Maison, 
Nucl. Phys, {\bf B383}, 357 (1992);  K. Lee, V.P. Nair, and E.J. Weinberg,
Phys. Rev. {\bf D45}, 2751 (1992).
\bibitem{cho} Kyoungtae Kimm, S.H. Oh, J.H. Yoon, and Y.M. Cho, to be published.


\end{thebibliography}
\end{document}